\documentstyle[amssymb,psfig]{mn}

\title[Mass segregation in rich LMC clusters III.]{Mass segregation in
rich clusters in the Large Magellanic Cloud -- III.  Implications for
the initial mass function}

\author[R.  de Grijs et al.]{R.  de Grijs,$^1$\thanks{E-mail:
grijs@ast.cam.ac.uk} G.F.  Gilmore,$^1$ A.D. Mackey,$^1$ M.I.  Wilkinson,$^1$
S.F. Beaulieu,$^2$ 
\newauthor
R.A.  Johnson,$^3$ and B.X. Santiago$^4$ \\
$^1$ Institute of Astronomy, University of Cambridge, Madingley Road,
Cambridge CB3 0HA \\
$^2$ Department of Physics and Astronomy, University of Victoria, 3800
Finnerty Road, Victoria, BC, V8P 1A1, Canada \\
$^3$ European Southern Observatory, Casilla 19001, Santiago 19, Chile \\
$^4$ Universidade Federal do Rio Grande do Sul, Instituto de F\'\i sica,
91501-970 Porto Alegre, RS Brazil
}

\date{Accepted ---. Received ---; in original form ---.}

\pubyear{2002}

\begin{document}
\maketitle

\begin{abstract}
The distribution of core radii of rich clusters in the Large Magellanic
Cloud (LMC) systematically increases in both upper limit and spread with
increasing cluster age.  Cluster-to-cluster variations in the stellar
initial mass function (IMF) have been suggested as an explanation.  We
discuss the implications of the observed degree of mass segregation in
our sample clusters for the shape of the initial mass function.  \\
Our results are based on {\sl Hubble Space Telescope}/WFPC2 observations
of six rich star clusters in the LMC, selected to include three pairs of
clusters of similar age, metallicity, and distance from the LMC centre,
and exhibiting a large spread in core radii between the clusters in each
pair.  \\
All clusters show clear evidence of mass segregation: (i) their
luminosity function slopes steepen with increasing cluster radius, and
(ii) the brighter stars are characterized by smaller core radii.  {\it
For all sample clusters}, both the slope of the luminosity function in
the cluster centres and the degree of mass segregation are similar to
each other, within observational errors of a few tenths of power-law
slope fits to the data.  This implies that their {\it initial} mass
functions must have been very similar, down to $\sim 0.8 - 1.0
M_\odot$. \\
We therefore rule out variations in the IMF of the individual sample
clusters as the main driver of the increasing spread of cluster core
radii with cluster age. 
\end{abstract}


\section{Introduction}
\label{intro.sec}

The Large Magellanic Cloud (LMC) contains massive star clusters at all
stages of their evolution, exhibiting a wide range of intrinsic physical
properties.  The focus of this paper is a detailed comparison among the
stellar populations in six rich LMC star clusters, which were chosen in
three pairs of similar age, mass, metallicity, and distance from the
centre of the LMC, but exhibiting a large range in core radii.  We have
chosen pairs of clusters with very different core radii at the same age
to test directly if variations in the initial mass function (IMF) are
the cause of the systematic increase in both the upper limit and spread
of the cluster core radii with increasing age seen in the rich clusters
in the Magellanic Clouds (e.g., Mackey \& Gilmore 2002 and references
therein). 

\subsection{The distribution of LMC cluster core radii}

\begin{figure}
\psfig{figure=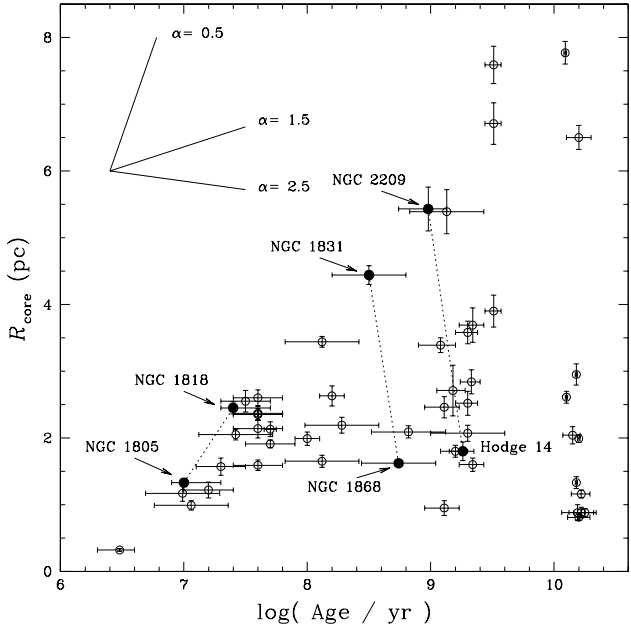,width=9cm}
\caption{\label{lmcclusters.fig}Distribution of core radius versus age
for all LMC clusters in the sample of Mackey \& Gilmore (2002).  The
clusters observed as part of our {\sl HST} programme GO-7307 are
indicated; pairs of our sample clusters spanning a large range of core
radii at (roughly) similar age are connected by dotted lines.  The solid
lines indicate the expected core evolution generated by an IMF with
slope $\alpha$.}
\end{figure}

In Fig.  \ref{lmcclusters.fig}, we show the distribution of cluster core
radii as a function of age in the LMC, using the most recent
determination of these properties by Mackey \& Gilmore (2002), based on
a randomly selected sample of 53 LMC clusters observed with the {\sl
Hubble Space Telescope (HST)}.  These authors confirm the observational
trend that the upper limits of the core radii systematically increase
with cluster age, as previously discussed by Elson, Freeman \& Lauer
(1989b), Elson (1991, 1992), and van den Bergh (1994), based on smaller
cluster samples observed from the ground.  This trend reflects true
physical evolution of the LMC cluster population, with some clusters
experiencing little or no core expansion, while others undergo
large-scale expansion due to some unknown process. 

One possible explanation is cluster-to-cluster variations in the IMF
(e.g., Elson et al.  1989b), and therefore different expansion rates of
the clusters due to varying mass loss rates of the evolving stellar
population (Chernoff \& Weinberg 1990).  However, the IMF slopes
required to explain the largest core radii are too flat to allow these
clusters to survive stellar mass loss beyond several $10^7$ yr (Elson
1991, Mackey \& Gilmore 2002), while an increasing body of evidence
points towards the universality of the IMF (see Gilmore 2001 for a
review). 

Alternative explanations for generating the largest core radii include
the dynamical effects of the binary stellar population in the cluster,
the merger of binary pairs of clusters (e.g., de Oliveira, Bica \&
Dottori 2000), and expansion due to tidal forces. 

We will evaluate the observational evidence in terms of these core
expansion mechanisms in Section \ref{tidal.sec}. 

\subsection{Effects of mass segregation}
\label{mseffects.sec}

Over the lifetime of a star cluster, encounters between its member stars
gradually lead to an increased degree of energy equipartition throughout
the cluster.  The most significant consequence of this process is that
the higher-mass cluster stars gradually sink towards the cluster centre
and in the process transfer their kinetic energy to the more numerous
lower-mass stellar component, thus leading to mass segregation.

The time-scale on which a cluster will have lost all traces of its
initial conditions is, to first order, well-represented by its
characteristic (half-mass) relaxation time, $t_{\rm r,h}$.  The
relaxation time-scale of a specific stellar species is directly related
to its mean mass.  Thus, significant mass segregation among the most
massive stars in the cluster core, occurs on the local, central
relaxation time-scale (comparable to just a few crossing times,
depending on the stellar mass, see Bonnell \& Davies 1998), whereas a
time-scale $\sim t_{\rm r,h}$ is required to affect a large fraction of
the cluster mass. 

However, the time-scale for a cluster to lose all traces of its initial
conditions also depends, among other factors, on (i) the smoothness of
its gravitational potential or, equivalently, the number of stars
(Bonnell \& Davies 1998); (ii) the degree of energy equipartition
reached (e.g., Hunter et al.  1995); and (iii) the slope of the mass
function (MF; e.g., Lightman \& Shapiro 1978, Inagaki \& Saslaw 1985,
Pryor, Smith \& McClure 1986, Sosin 1997). 

As the dynamical evolution of a cluster progresses, low-mass stars will,
on average, attain larger orbits than the cluster's higher-mass stars,
and the low-mass stars will thus spend most of their time in the
cluster's outer regions, at the extremes of their orbits.  For this
reason alone, we would not expect to achieve global equipartition in a
cluster (e.g., Inagaki \& Saslaw 1985).  In these outer parts, the
cluster's gravitational potential is weaker and constantly changing due
to the ongoing redistribution of mass (Chernoff \& Weinberg 1990), and
it is more easily affected by the tidal field in which the cluster
resides. 

In these circumstances, two effects will enhance the mass segregation
signatures observed in old, evolved clusters, (i) evaporation and
ejection across the cluster's tidal boundary of (preferentially)
low-mass stars, because of their higher velocity dispersion and number
density (Chernoff \& Weinberg 1990, Giersz \& Heggie 1997), and (ii)
tidal stripping by the external gravitational field of the low-mass
stars sent to the cluster's outer regions by the relaxation process in
the inner regions. 

We will discuss the effects of the tidal field on a cluster's degree of
mass segregation in relation to its size in Section \ref{tidal.sec}.  In
Section \ref{observations.sec}, we present our sample of six rich LMC
clusters, for which we analyse the degree of mass segregation attained
over their lifetimes in Section \ref{masssegr.sec}, based on the
clusters' luminosity functions (LFs) derived in Section \ref{mass.sec}. 

\section{Sample selection, observations and data reduction}
\label{observations.sec}

\subsection{Our LMC cluster sample}

As part of {\sl HST} GO programme 7307, we obtained {\sl WFPC2} imaging
observations of the rich LMC clusters in Table \ref{sample.tab}, where
we have also included a few of their basic properties.  Their location
in the (log(age) vs.  $R_{\rm core}$) diagram is indicated in Fig. 
\ref{lmcclusters.fig}.  For a full overview of the clusters' physical
parameters, we refer the reader to {\tt
http://www.ast.cam.ac.uk/STELLARPOPS/LMCdatabase/}. 

The clusters in our sample are among the richest in the LMC, and have
masses $\sim 10^4 M_\odot$.  Half-mass radii are typically $\sim 50$
arcsec, and maximum radii $\sim 200$ arcsec.  (At the distance of the
LMC, $\sim 52$ kpc, 4 arcsec $\approx 1$ pc).  Crossing times at the
half-mass radius are $\sim 10^7$ years, and characteristic two-body
relaxation times are $\sim 10^6 - 10^8$ years in the cluster core and
$\sim 10^9$ years at the half-mass radius (Elson, Fall \& Freeman 1989a,
see also de Grijs et al.  2002a,b, hereafter Papers I,II, for NGC 1805
and NGC 1818).  Our clusters are chosen with ages spanning this range,
and should thus resolve the evolutionary processes that operate on each
time-scale.  They are paired in age to help discriminate between trends
and coincidences (eg., in the [initial] MF, see Section
\ref{masssegr.sec}), and each pair is at a similar distance (and if
possible in a similar direction) from the centre of the LMC (both the
geometrical centre [Bica et al.  1996], and the dynamical, rotation
centre [see Westerlund 1990]), to minimise any differential effects of
the tidal field of the LMC on the cluster's evolution.  They are also at
the greatest possible distance from the LMC centre, where the effects of
the tidal field are smaller, and stellar backgrounds are sparser.  The
total radial range occupied by our sample clusters ranges from about 3.5
to 5.5 degrees for the entire sample (out of the full radial range
occupied by the LMC cluster sample from $< 1$ to $\sim 15$ degrees),
with differences between the two clusters in each pair of less than a
degree.  Care was taken to avoid clusters exhibiting post-core-collapse
(PCC) characteristics. 

\begin{table*}
\caption[ ]{\label{sample.tab}Our LMC cluster sample}
{\scriptsize
\begin{center}
\begin{tabular}{ccccclccccc}
\hline
\multicolumn{1}{c}{Cluster} & \multicolumn{1}{c}{log(age) [yr]} &
\multicolumn{1}{c}{Ref.} & \multicolumn{1}{c}{[Fe/H] (dex)} &
\multicolumn{1}{c}{Ref.} & \multicolumn{1}{c}{log($m/M_\odot$)} &
\multicolumn{1}{c}{Ref.} & \multicolumn{1}{c}{$R_{\rm core}$ (pc)$^a$} &
\multicolumn{1}{c}{Ref.} & \multicolumn{1}{c}{$D_{\rm LMC}$ (deg)$^b$} & 
\multicolumn{1}{c}{Ref.} \\
\hline
NGC 1805 & $7.0 \pm 0.05$ & 4,14 & $-0.4 - 0.0$ & 10 & $3.52 \pm 0.13$        & 11 & $1.33 \pm 0.06$ & 1,11  & $3.86 - 4.00$ & 12 \\
NGC 1818 & $\sim 7.4$     & 2,4  & $-0.4 - 0.0$ & 10 & $4.13^{+0.15}_{-0.14}$ & 11 & $2.45 \pm 0.09$ & 1,11  & $3.47 - 3.61$ & 12 \\
NGC 1831 & $\sim 8.6$     & 3,9  & $-0.35$    & 3  & $4.81 \pm 0.13$        & 11 & $4.44 \pm 0.14$ & 11    & $4.82 - 4.85$ & 12 \\
NGC 1868 & $8.70 \pm 0.2$ & 6,12,13 & $-0.50$ & 13 & $4.53 \pm 0.10$        & 11 & $1.62 \pm 0.05$ & 11    & $5.57 - 5.47$ & 12 \\
NGC 2209 & $\sim 9.0$     & 8,9,12 & $-1.1$ & 5,7 & $5.03^{+0.36}_{-0.6}$ & 11 & $5.43 \pm 0.33$ & 11      & $5.48 - 5.43$ & 12 \\
Hodge 14 & $\sim 9.2$     & 6,8,13 & $-0.66 \pm 0.2$ & 13 & $4.33^{+0.34}_{-0.28}$ & 11 & $1.80 \pm 0.14$ & 11 & $4.19 - 4.37$ & 12 \\
\hline
\end{tabular}
\end{center}
\flushleft
Notes: $^a$ Based on distance moduli determined by Castro et al.  (2001);
$^b$ Distance from the LMC centre in degrees, w.r.t.  the optical,
geometrical centre (Bica et al.  1996), and the dynamical, rotation
centre (see Westerlund 1990).  \\
References: 1, paper I; 2, paper II; 3, Bonatto et al.  (1995); 4,
Cassatella et al.  (1996); 5, Chiosi et al.  (1986); 6, Elson et al. 
(1989a); 7, Frogel et al.  (1990); 8, Geisler et al.  (1997); 9, Girardi
et al.  (1995); 10, Johnson et al.  (2001); 11, Mackey \& Gilmore
(2002); 12, Meurer et al.  (1990); 13, Olszewski et al.  (1991); 14,
Santos et al.  (1995)
} 
\end{table*}

\subsection{{\sl HST/WFPC2} Observations}

In this section, we will give a brief overview of the available {\sl
WFPC2} data for the star clusters in our sample.  This paper builds on
the preparatory research by Santiago et al.  (2001) for the entire
sample, and is an extension of Papers I and II, which focused on mass
segregation in the youngest pair of LMC clusters in our sample, NGC 1805
and NGC 1818. 

We obtained {\sl WFPC2} exposures through the F555W and F814W filters
(roughly corresponding to the Johnson-Cousins {\it V} and {\it I}
filters, respectively) for each cluster, with the PC centred on both the
cluster centre, and on its half-mass radius.  Following Santiago et al. 
(2001), we will refer to these two sets of exposures as our CEN and HALF
fields, respectively.  For the CEN fields, we obtained both deep and
shallow images.  Exposure times for the former were 140s and 300s,
respectively, for each individual image in F555W and F814W, while for
the latter exposure times of 5s and 20s were used for the F555W and
F814W filters, respectively.  The shallow exposures were intended to
obtain aperture photometry for the brightest stars in the cluster
centres, which are saturated in the deeper exposures.  For the HALF
field, we obtained deep observations with a total exposure time of 2500s
through each filter.  At each position, for each set of deep and shallow
exposures, and through both filters, we imaged our clusters in sets of 3
observations, to facilitate the removal of cosmic rays.  The
observations obtained for NGC 1805 and NGC 1818 were described in detail
in Paper I; in Table \ref{observations.tab} we present an overview of
the observations obtained for the remaining LMC clusters in our sample. 

\begin{table*}
\caption[ ]{\label{observations.tab}Overview of the additional {\sl WFPC2} observations
}
{\scriptsize
\begin{center}
\begin{tabular}{lllrllrc}
\hline
\multicolumn{1}{c}{Object} & \multicolumn{1}{c}{Field} &
\multicolumn{1}{c}{Filter} & \multicolumn{1}{c}{Exposure} &
\multicolumn{1}{c}{RA$^a$} & \multicolumn{1}{c}{Dec$^a$} &
\multicolumn{1}{c}{Position} & \multicolumn{1}{c}{Date (UT)} \\
& & & time (s) & \multicolumn{2}{c}{(J2000)} & angle ($^\circ$)$^b$ &
(dd/mm/yyyy) \\
\hline
NGC 1831 & CEN  & F555W & 3x5    & 05:06:16.502 & $-$64:55:06.391 &  $-$90.86 & 25/07/1998 \\
         &      &       & 3x140  & \\
         &      & F814W & 3x20   & & & & 25/07/1998 \\
         &      &       & 3x300  & \\
         & HALF & F555W & 2x800  & 05:06:08.846 & $-$-64:55:05.481 & $-$142.01 & 29/05/1998 \\
         &      &       & 900    & \\
         &      & F814W & 3x800  & & & & 29/05/1998 \\
         &      &       & 900    & \\
NGC 1868 & CEN  & F555W & 3x5    & 05:14:36.061 & $-$63:57:16.460 & 22.39 & 12/11/1998 \\
         &      &       & 3x140  & \\
         &      & F814W & 3x20   & & & & 12/11/1998 \\
         &      &       & 3x300  & \\
         & HALF & F555W & 2x800  & 05:14:36.025 & $-$63:57:34.112 & $-$153.09 & 20/05/1998 \\
         &      &       & 900    & \\
         &      & F814W & 3x800  & & & & 24/05/1998 \\
         &      &       & 900    & \\
NGC 2209 & CEN  & F555W & 3x5    & 06:08:35.135 & $-$73:50:12.084 & 145.97 & 29/03/1998 \\
         &      &       & 3x140  & \\
         &      & F814W & 3x20   & & & & 29/03/1998 \\
         &      &       & 3x300  & \\
         & HALF & F555W & 2x800  & 06:08:37.047 & $-$73:49:07.330 & 177.89 & 06/05/1998 \\
         &      &       & 900    & \\
         &      & F814W & 3x800  & & & & 05/05/1998 \\
         &      &       & 900    & \\
Hodge 14 & CEN  & F555W & 3x5    & 05:28:37.884 & $-$73:37:50.214 & 153.32 & 31/03/1998 \\
         &      &       & 3x140  & \\
         &      & F814W & 3x20   & & & & 31/03/1998 \\
         &      &       & 3x300  & \\
         & HALF & F555W & 2x800  & 05:28:33.458 & $-$73:38:08.567 & 164.91 & 04/05/1998 \\
         &      &       & 900    & \\
         &      & F814W & 3x800  & & & & 04/05/1998 \\
         &      &       & 900    & \\
\hline
\end{tabular}
\end{center}
\flushleft
$^a$ centre of the PC; $^b$ East w.r.t. North.
}
\end{table*}

The pixel size of the WF and PC chips is 0.097 and 0.0455 arcsec,
respectively, with a total combined field of view of roughly 4850
arcsec$^2$ for the entire {\sl WFPC2} detector. 

\subsection{Initial data processing}
\label{processing.sec}

To obtain the clusters' luminosity functions (LFs) used in this paper we
followed identical procedures as discussed in Paper I, based on the
pipeline image reduction and recalibration of the {\sl WFPC2} images
using the updated and corrected on-orbit flat fields and related
reference files most appropriate for our observations. 

\begin{figure*}
\hspace{1.2cm}
\psfig{figure=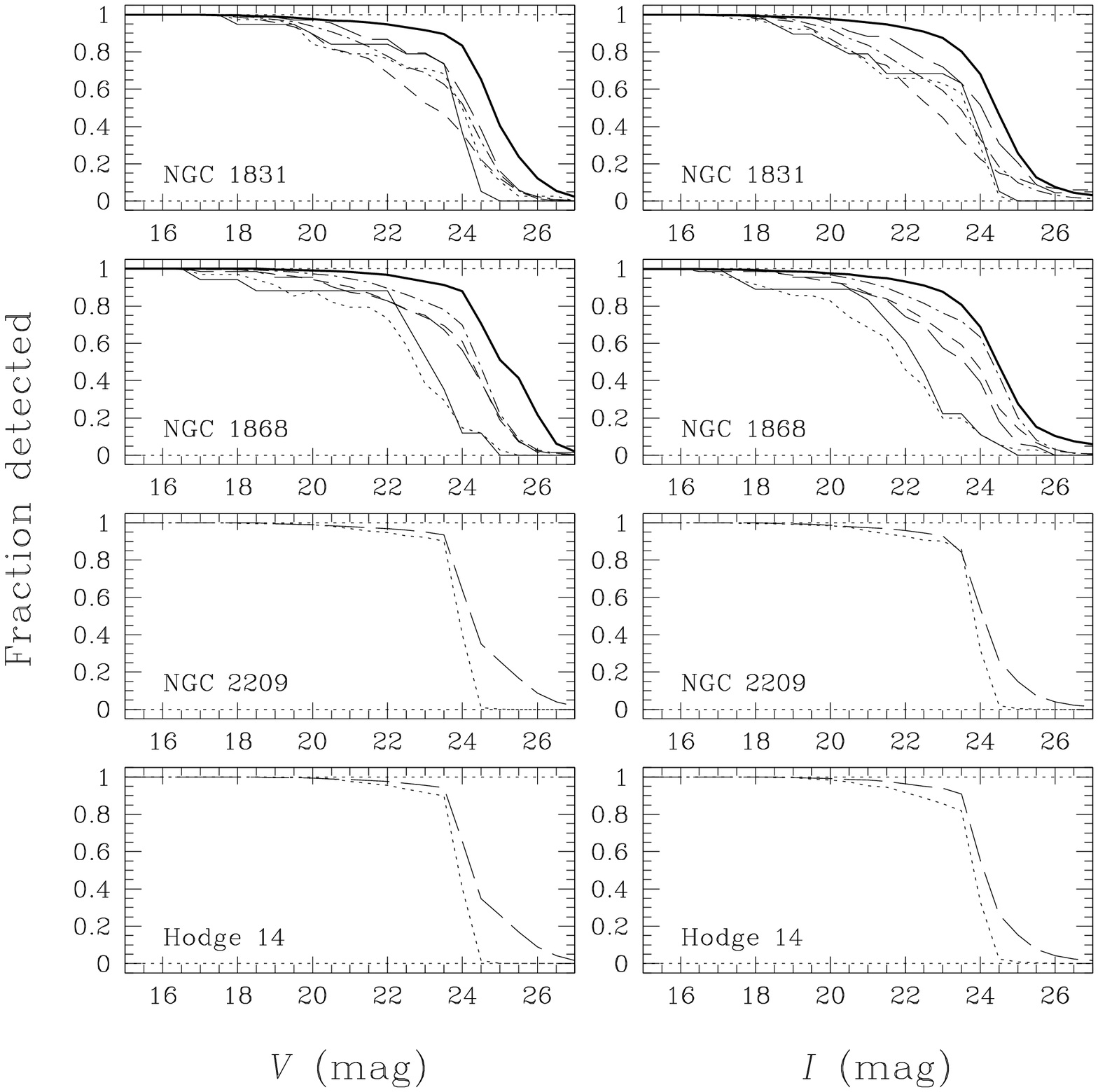,width=15cm}
\caption{\label{compl.fig}Completeness curves for NGC 1831, NGC 1868,
NGC 2209 and Hodge 14.  The different line styles refer to different
annuli.  {\it (i)} NGC 1831 and NGC 1868: thin solid -- $r \le 3.6$
arcsec; dotted -- $3.6 < r \le 7.2$ arcsec; long dashes -- $7.2 < r \le
18$ arcsec; short dashes -- $18 < r \le 36$ arcsec; dash-dotted -- $36 <
r \le 54$ arcsec; thick solid -- $r > 54$ arcsec.  {\it (ii)} NGC 2209
and Hodge 14: dotted -- $r \le 18$ arcsec; long dashes -- $r > 18$
arcsec.}
\end{figure*}

As in Paper I, owing to the significant stellar density gradient across
the cluster fields, completeness corrections are a strong function of
position within a cluster.  Therefore, we computed completeness
corrections for all observations in circular annuli around the centre of
each cluster, for both the PC and the WF fields, located at intervals
between the centre and 3.6 arcsec, $3.6-7.2$ arcsec, $7.2-18.0$ arcsec,
$18.0-36.0$ arcsec, $36.0-54.0$ arcsec and at radii $\ge 54.0$ arcsec
for NGC 1831 and NGC 1868.  The much sparser appearance of NGC 2209 and
Hodge 14 allowed us to sample their completeness functions using only
two radial ranges, for radii smaller and greater than 18 arcsec,
respectively.  The results of this exercise, based on the long CEN and
HALF exposures, are shown in Fig.  \ref{compl.fig}.  These completeness
curves were corrected for the effects of blending or superposition of
multiple randomly placed artificial stars as well as for the
superposition of artificial stars on genuine objects (see Paper I for a
full discussion).  The progressive increase in completeness fraction
with radius for a given source brightness, in particular for NGC 1831
and NGC 1868, clearly illustrates the potentially serious effects of
crowding in the inner regions of the clusters.  In the analysis
performed in this paper, we only include those ranges of the stellar LF
where the completeness fraction is in excess of 50 per cent. 

\section{Luminosity functions}
\label{mass.sec}

In the remainder of this paper, we will examine the dependence of the
shape and slope of the stellar LF on position within the clusters. 
Since there is a one-to-one correlation between a cluster's LF and its
MF, we will use these terms interchangeably.  However, in view of the
uncertainties involved in the conversion of luminosities to masses (see
Paper II for a detailed discussion), in this paper we will {\it only}
use the LFs to reach our conclusions on the effects of mass segregation. 
This approach is therefore less model-dependent and leads to identical
results, without having to keep in mind the large systematic
uncertainties inherent to any luminosity-to-mass conversion (see Paper
II). 

Where the full 2-dimensional CMDs were used in the literature to infer
the presence and the effects of mass segregation, this was mostly based
on differences in the concentration of specific stellar types, most
often main-sequence and giant branch stars.  However, the cluster
stars in our young and intermediate-age clusters start to saturate at
the faint end of the red giant branch, so this approach is not feasible. 
In fact, with the exception of a handful of the brightest stars, in our
cluster sample we are limited to the analysis of main-sequence stars
between the main-sequence turn-off (MSTO) and the 50 per cent
completeness limit. 

First, we need to correct the observed stellar LFs in the CEN and HALF
fields for the contribution of stars from the LMC background in these
fields.  We used the background field LFs described in Paper I (see also
Castro et al.  2001) for this purpose.  Thus, when we refer to the
cluster LFs in the remainder of this paper, this applies to the
background and completeness-corrected LFs.  Foreground stars are not a
source of confusion in the case of our LMC clusters, for $V \lesssim
23$, as already shown in Paper I.  Since this is consistent with the
standard Milky Way star count models (e.g., Ratnatunga \& Bahcall 1985)
and supported by the appearance of their CMDs, we extrapolate this
result to the other four star clusters included in the present study. 

\begin{figure*}
\hspace{1.2cm}
\psfig{figure=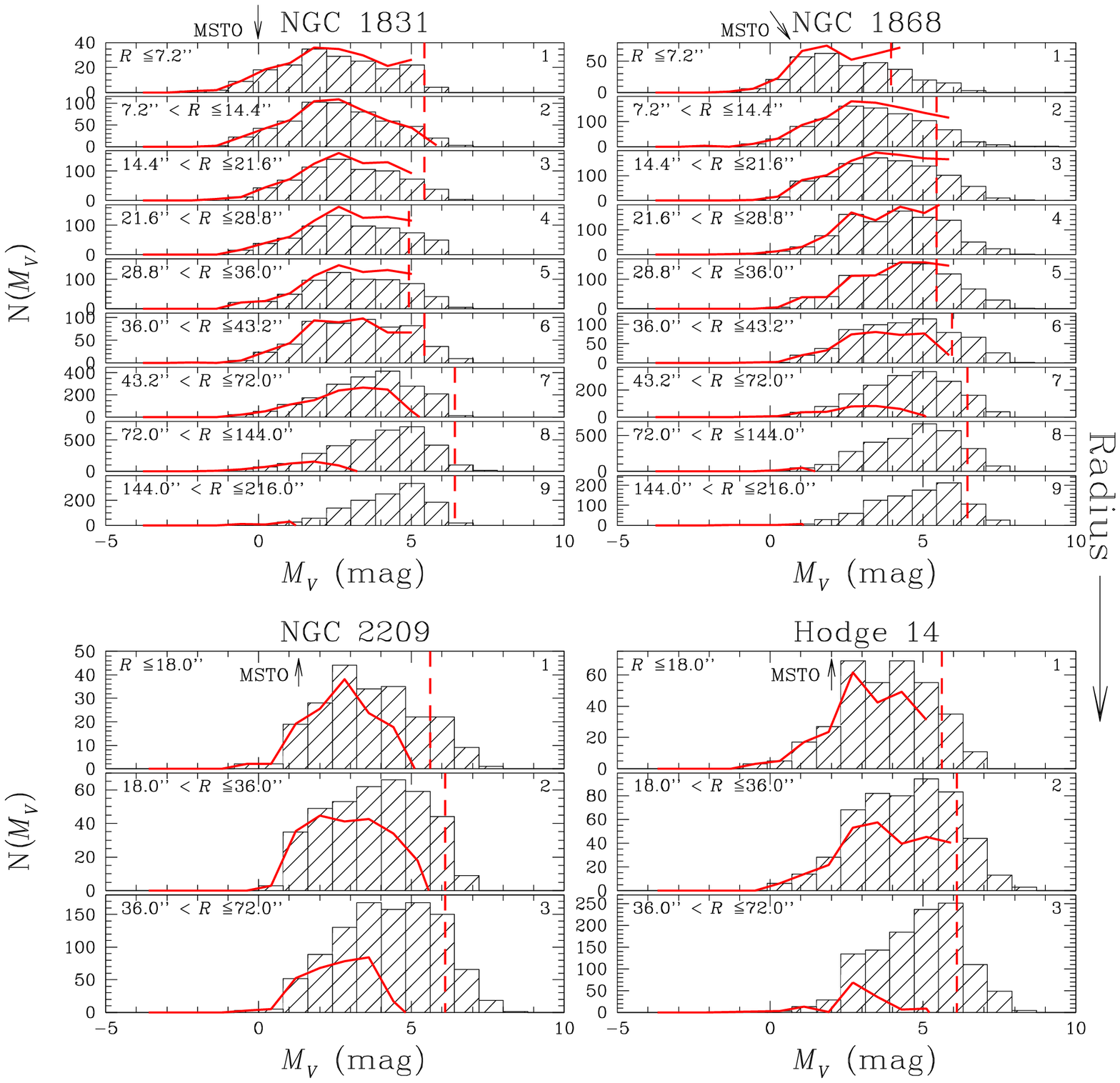,width=15cm}
\caption{\label{clfdata.fig}Observational total LFs in annuli at
increasingly large radii from the cluster centres (histograms).  The
thick solid lines are the actual cluster star distributions, after
correction for the background field star contribution and the effects of
incompleteness; the 50 per cent completeness limits are indicated by the
vertical dashed lines.  The MSTO magnitudes are indicated by the arrows.}
\end{figure*}

Fig.  \ref{clfdata.fig} shows the distribution of stellar magnitudes as
a function of distance from the cluster centres.  The shaded histograms
represent the total number of stars in our final source lists, not
corrected for incompleteness, area covered or background star
contamination; the thick solid lines are the actual cluster star
distributions, obtained by subtracting the background contribution
expected in the area covered by each annulus from the observed total LFs
and subsequently correcting for incompleteness effects.  The 50 per cent
completeness limits in each annulus are indicated by the vertical dashed
lines through the centres of the last magnitude bin above this limit. 

\begin{figure*}
\hspace{1.2cm}
\psfig{figure=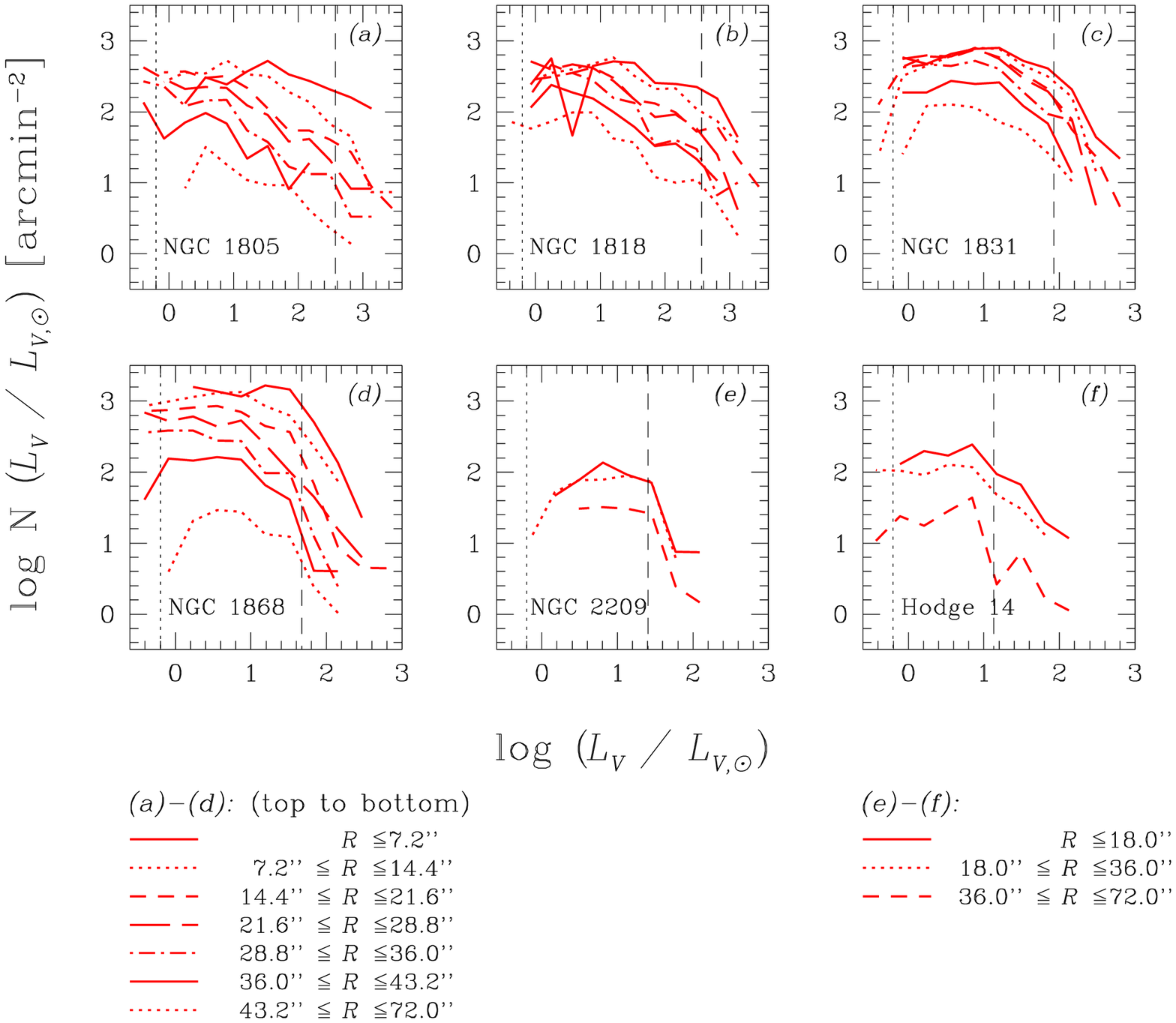,width=15cm}
\vspace*{-1.5cm}
\caption{\label{allclfs.fig}Corrected LFs: comparison of annular LFs
from the inner 7 annuli shown in Fig.  \ref{clfdata.fig} and in Paper I
for NGC 1805 and NGC 1818, normalised to 1 arcmin$^2$ area coverage. 
For reasons of clarity, we have omitted the vertical error bars.  The
approximate MSTO luminosities are indicated by the vertical dashed
lines; the dotted lines represent the faintest luminosity range used for
fitting the LF slopes (see text).}
\end{figure*}

In Fig.  \ref{allclfs.fig} we show all annular cluster LFs out to $R =
72.0$ arcsec, corrected for the effects of incompleteness (as a function
of radial distance from the cluster centres), background contamination
and for the sampling area covered by each (partial) annulus, for all of
our sample clusters.  In this representation, the radial dependence of
the cluster LFs is more easily visible than in Fig.  \ref{clfdata.fig}. 

\section{The slope of the luminosity function}
\label{masssegr.sec}

We subsequently determined the LF slopes, assuming a simple power-law
dependence for the number of stars of a given luminosity, i.e., $N(L)
\propto L^{-\alpha}$, where $\alpha$ is the LF slope.  We realise,
however, that the inner LFs in Fig.  \ref{allclfs.fig} show clear maxima
inside our fitting ranges in most cases, and that the overall cluster
LFs are clearly {\it not} linear. 

\begin{figure*}
\hspace{1.2cm}
\psfig{figure=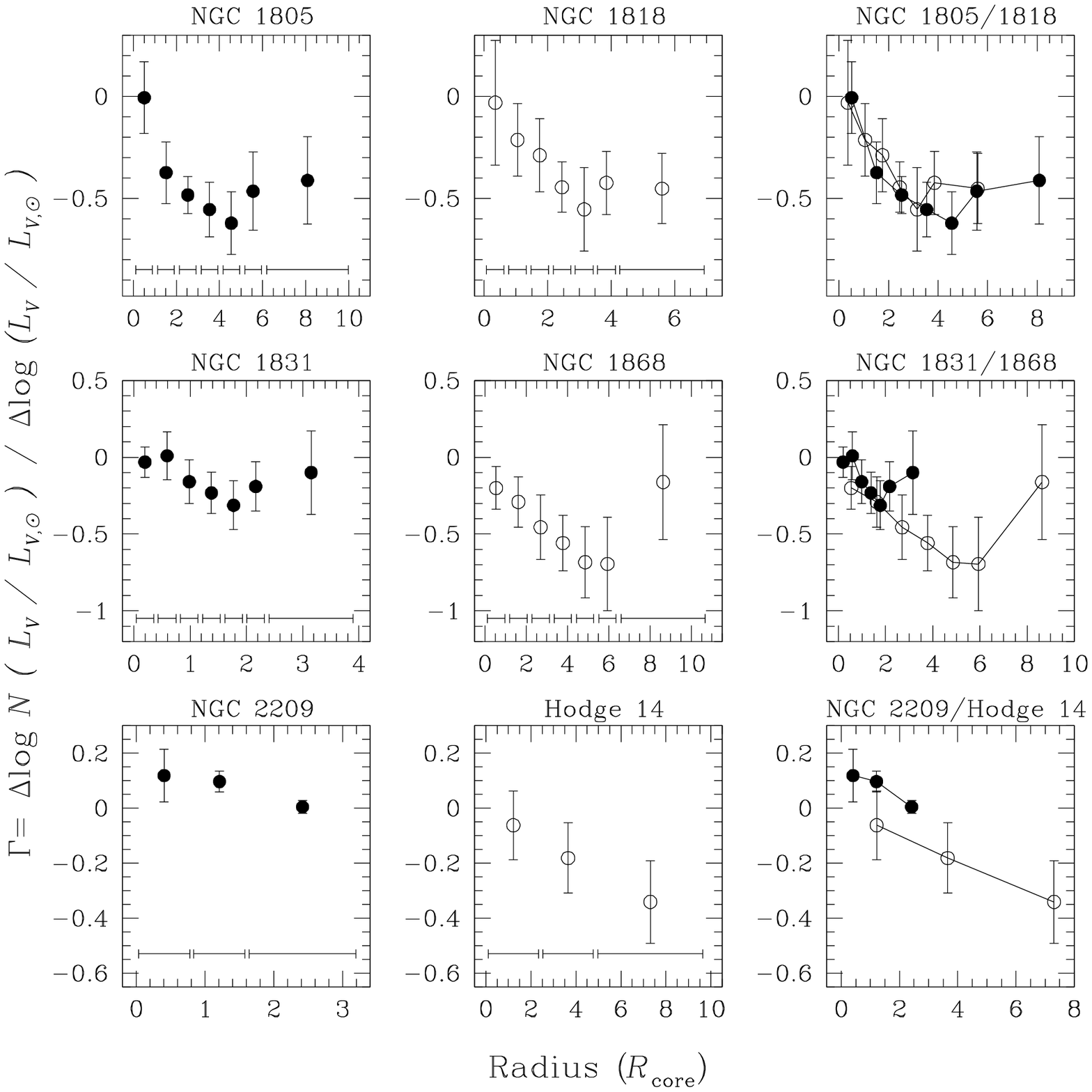,width=15cm}
\caption{\label{comparelf.fig}LF slope as a function of cluster radius,
expressed in units of their core radii, and comparison between the LF
slopes of the two clusters in each pair.  The slopes were determined
over the maximum available luminosity range, from the MSTO to the 50 per
cent completeness limit, for each pair of clusters of similar age (see
Fig.  \ref{allclfs.fig}): {\it (i)} NGC 1805 and NGC 1818: $-1.60 \le
M_V \le 5.10 \; (2.57 \ge \log L_V / L_{V,\odot} \ge -0.11)$; {\it (ii)}
NGC 1831 and NGC 1868: $0.65 \le M_V \le 5.10 \; (1.67 \ge \log L_V /
L_{V,\odot} \ge -0.11)$; {\it (iii)} NGC 2209 and Hodge 14: $2.00 \le
M_V \le 5.10 \; (1.13 \ge \log L_V / L_{V,\odot} \ge -0.11)$.  The
radial ranges over which the LF slopes were determined are shown by
horizontal bars at the bottom of each panel.}
\end{figure*}

Despite this, a comparison of LF slopes obtained using power-law fits
over identical luminosity ranges is still valuable to quantify the
radial dependence of the cluster LFs.  We chose to use fitting ranges in
luminosity that covered the maximum overlap among our annular LFs
between the clusters in each pair, in order to minimise the effects of
small-scale statistical fluctuations in the LFs.  The results are shown
in Fig.  \ref{comparelf.fig}.  For NGC 1805 and NGC 1818, we used the
ranges $-1.60 \le M_V \le 5.10 \; (2.57 \ge \log L_V / L_{V,\odot} \ge
-0.11)$; for NGC 1831 and NGC 1868, $0.65 \le M_V \le 5.10 \; (1.67 \ge
\log L_V / L_{V,\odot} \ge -0.11)$; and for NGC 2209 and Hodge 14, $2.00
\le M_V \le 5.10 \; (1.13 \ge \log L_V / L_{V,\odot} \ge -0.11)$.  These
ranges correspond to the luminosity range between the faintest MSTO
magnitude in each pair and the corresponding 50 per cent completeness
limit.  The radial ranges to which the data points apply are indicated
by small bars at the bottom of each panel. 

In all of our sample clusters the LF slopes clearly steepen with
increasing cluster radius.  This corresponds to clear mass segregation,
the amount of which is sensitively dependent on the luminosity-to-mass
conversion used (see Paper II).  Although the trend towards steeper LFs
with increasing radius is clear, the associated error bars are large. 
They are dominated by the non-linearity of the annular LFs and
point-to-point variations. 

Nevertheless, it is apparent from the comparison of the dependence of
the LF slope as a function of radius (expressed in units of their core
radii) between the clusters in each pair (right-hand panels of Fig. 
\ref{comparelf.fig}), that, within the uncertainties, this dependence is
identical for both clusters in a given age pair. 

\begin{figure}
\psfig{figure=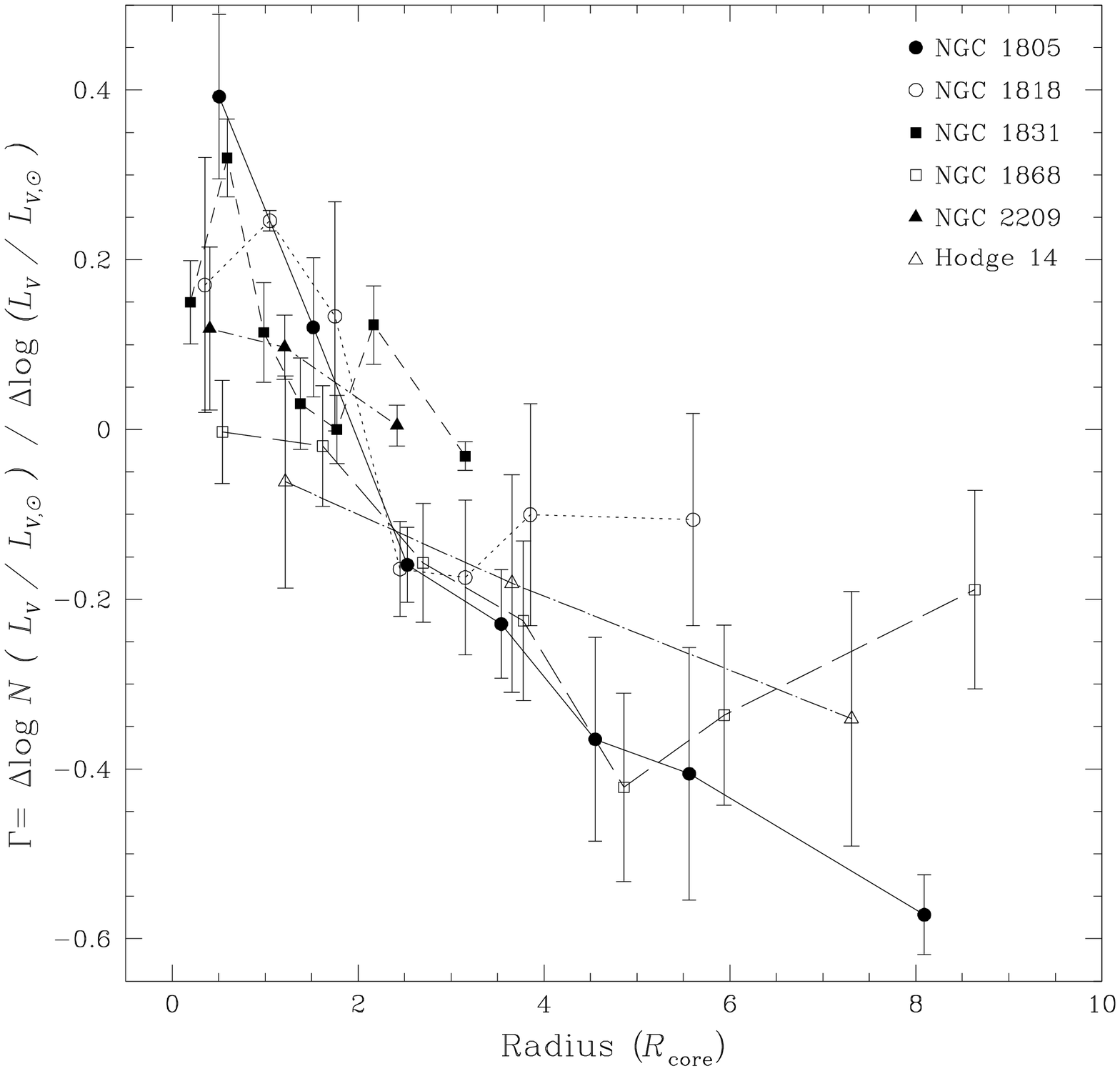,width=9cm}
\caption{\label{compareall.fig}Comparison of the LF slopes for all
clusters in our LMC sample. The slopes were determined over the greatest
luminosity range in common among all clusters, $2.00 \le M_V \le 5.10 \;
(1.13 \ge \log L_V / L_{V,\odot} \ge -0.11)$, limited by the MSTO
location of the oldest sample clusters.}
\end{figure}

In Fig.  \ref{compareall.fig} we compare the dependence of the LF slope
on cluster radius for all of our sample clusters.  The LF slopes were
determined over the largest possible common luminosity range, $2.00 \le
M_V \le 5.10 \; (1.13 \ge \log L_V / L_{V,\odot} \ge -0.11)$, i.e., from
the MSTO of Hodge 14 to the 50 per cent completeness limit at $M_V
\simeq 5.1$.  Surprisingly, we find that {\it both} the central LF slope
{\it and} the degree of mass segregation of all sample clusters, as
indicated by the gradient of the LF slope with radius, are confined
within narrow ranges, at most spanning a $\sim (2-3) \sigma$ range in
parameter space.  This is a robust result, and is indeed rather
surprising in view of the large range in age (and therefore in dynamical
state), mass, metallicity, and structural parameters (core radii)
occupied by the ensemble of our sample clusters.  We will discuss the
implications of this result in more detail in Section \ref{tidal.sec}. 

\begin{figure*}
\hspace*{1.2cm}
\psfig{figure=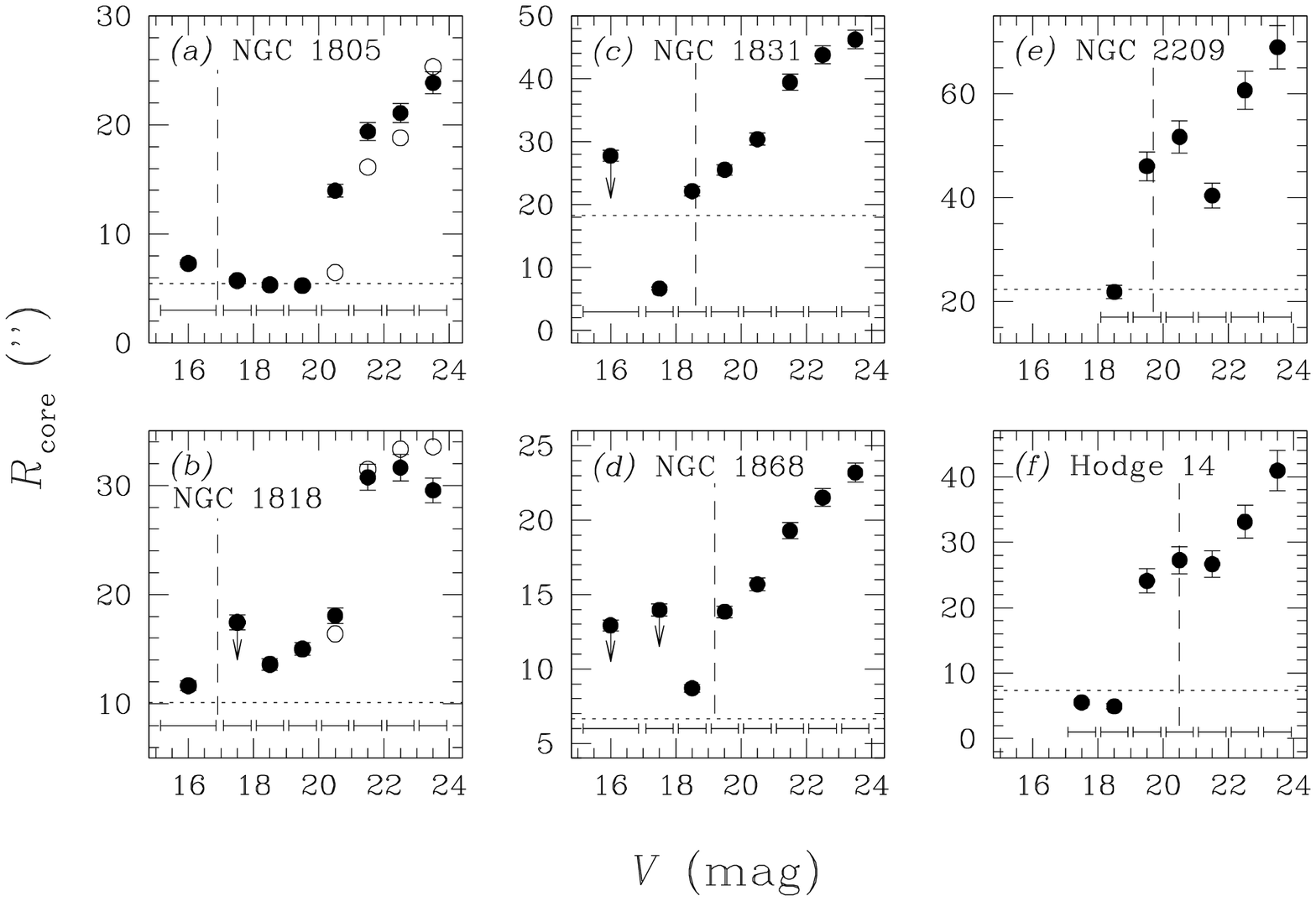,width=15cm}
\vspace*{-4.2cm}
\caption{\label{allcoreradii.fig}Core radii as a function of magnitude
(mass) for our cluster sample.  The filled circles are the core radii
after correction for the effects of (in)completeness, area covered by
the observations, and background stars; the open circles in panels {\it
(a)} and {\it (b)} are not background subtracted and serve to indicate
the uncertainties due to background correction.  We have also indicated
the mean cluster core radii, obtained from surface brightness profile
fits (dotted lines; Mackey \& Gilmore 2002).  The horizontal bars at the
bottom of the panels indicate the magnitude ranges used to obtain the
core radii; from bright to faint magnitudes, the centres of the
magnitude ranges correspond approximately to $\log (m / M_\odot) = 0.90,
0.75, 0.60, 0.44, 0.30, 0.16, 0.09$, and 0.00, respectively, the exact
value depending sensitively on the luminosity-to-mass conversion used
(see Paper II).  The data points with arrows indicate upper limits, as
explained in the text.  The vertical dashed lines indicate the MSTO in
each cluster.}
\end{figure*}

Finally, we determined the dependence of core radius on the adopted
luminosity (magnitude) range, as shown in Fig.  \ref{allcoreradii.fig}. 
Core radii were derived based on fits to stellar number counts --
corrected for the effects of incompleteness\footnote{We only used
magnitude (mass) ranges for which the completeness fractions were at
least 50 per cent.} and background contamination -- of the generalised
fitting function proposed by Elson, Fall \& Freeman (1987a), in the
linear regime:
\begin{equation}
\label{elson.eq}
\mu(r) = \mu_0 \Biggl( 1 + \Bigl( {r \over a} \Bigr)^2
\Biggr)^{-\gamma/2} ,
\end{equation}
where $\mu(r)$ and $\mu_0$ are the radial and central surface
brightness, respectively, $\gamma$ corresponds to the profile slope in
the outer regions of the cluster, and $R_{\rm core} \approx a
(2^{2/\gamma} - 1)^{1/2} \approx R_{\rm h}$.

For all of our sample clusters we clearly see the effects of mass
segregation, in the sense that the brighter stars below the MSTO
magnitude are increasingly concentrated towards the cluster centres
(i.e., they are characterized by smaller core radii).  In Table
\ref{coreradii.tab} we list, for stars below the MSTO, the core radii
for each magnitude range for the oldest four sample clusters (for NGC
1805 and NGC 1818 these numbers were published in Paper II). 

\begin{table*}
\caption[ ]{\label{coreradii.tab}Cluster core radii as a function of
mass, below the MSTO}
{\scriptsize
\begin{center}
\begin{tabular}{cccccccccc}
\hline
\hline
\multicolumn{1}{c}{Magnitude} & \multicolumn{1}{c}{$\log m/M_\odot$} &
\multicolumn{2}{c}{NGC 1831} & \multicolumn{2}{c}{NGC 1868} &
\multicolumn{2}{c}{NGC 2209} & \multicolumn{2}{c}{Hodge 14}\\
\multicolumn{1}{c}{range $(V)$} & (central) & \multicolumn{1}{c}{$('')$}
& \multicolumn{1}{c}{(pc)} & \multicolumn{1}{c}{$('')$} &
\multicolumn{1}{c}{(pc)} & \multicolumn{1}{c}{$('')$} &
\multicolumn{1}{c}{(pc)} & \multicolumn{1}{c}{$('')$} &
\multicolumn{1}{c}{(pc)} \\
\hline
$19.0 - 20.0$ & 0.44 & 25.54 &  6.44 & 13.85 & 3.44 & $\dots$ & $\dots$ & $\dots$ & $\dots$ \\
$20.0 - 21.0$ & 0.30 & 30.39 &  7.66 & 15.69 & 3.90 & 51.68   & 11.93   & 44.89   & 10.86 \\
$21.0 - 22.0$ & 0.16 & 39.44 &  9.94 & 19.30 & 4.80 & 40.39   &  9.32   & 50.91   & 12.32 \\
$22.0 - 23.0$ & 0.09 & 43.79 & 11.04 & 21.52 & 5.35 & 60.66   & 14.00   & 45.99   & 11.12 \\
$23.0 - 24.0$ & 0.00 & 46.22 & 11.65 & 23.20 & 5.77 & 68.94   & 15.91   & 49.90   & 12.07 \\
\hline
\end{tabular}
\end{center}
}
\end{table*}

The slightly larger scatter for NGC 1818, NGC 2209 and Hodge 14 is due
to the smaller number of stars in each magnitude bin compared to the
other clusters; for these three clusters the associated uncertainties
are determined by a combination of the scatter in the derived core radii
and background effects, while the uncertainties for the others are
dominated by the effects of background subtraction.  The upper limits
for the core radii determined from the distribution of the brightest
cluster stars ($V \lesssim 18$) in NGC 1818, NGC 1831 and NGC 1868 are
due to the combined effects of background corrections, small number
fluctuations, and the radial sampling of the stellar distributions. 

We have also indicated the core radii obtained from profile fits to the
overall surface brightness profiles of the clusters.  It is clear that
these are dominated by the mass-segregated high-mass (bright) stars.  It
is also encouraging to notice that the core radii determined
independently from the surface brightness profiles and those from the
brighter cluster stars are internally consistent. 

\subsection{Comparison with previously published results}
\label{comparison.sec}

\begin{figure*}
\hspace{1.2cm}
\psfig{figure=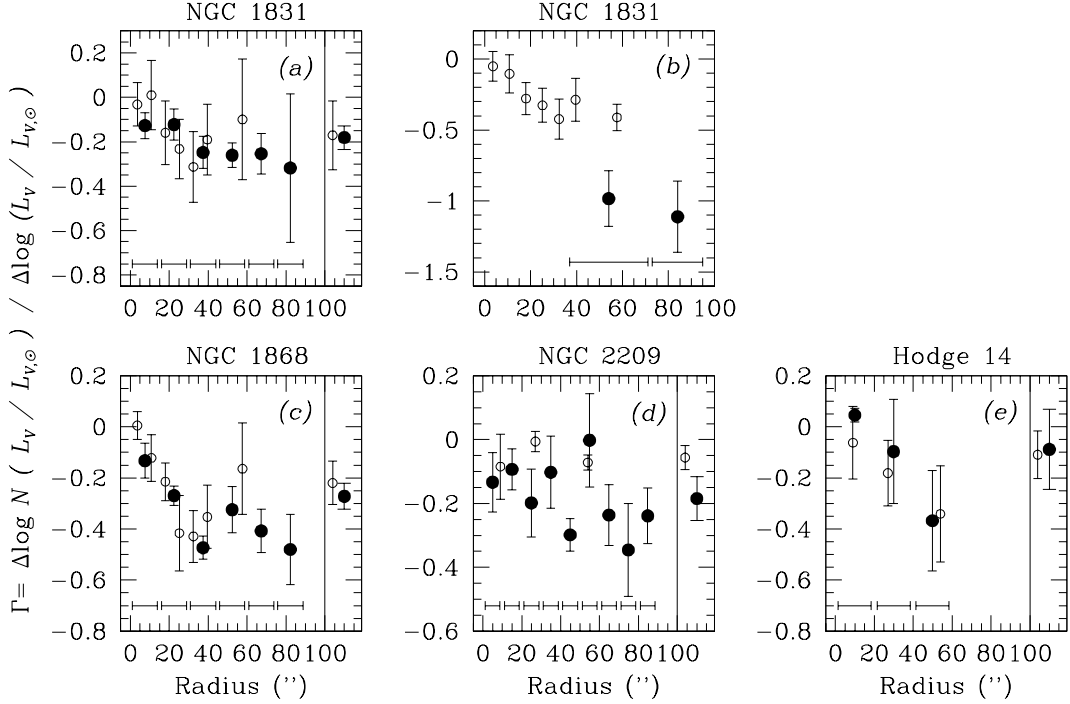,width=15cm}
\vspace*{-5cm}
\caption{\label{slopecf.fig}Comparison of our LF slopes with those
published in the literature.  (a), (c)--(e) Comparison with Santiago et
al.  (2001) after redetermination of the LF slopes using identical
absolute magnitude ranges for each sample, as described in the text. 
The right-hand subpanels show the global LF slopes for the clusters. 
(b) Comparison with Mateo (1988).  Filled bullets: literature data, open
circles: this paper.  The radial ranges used to obtain the literature
data are indicated by the horizontal bars at the bottom of each panel.}
\end{figure*}

In Paper I, we compared our LF slopes as a function of cluster radius
for NGC 1805 and NGC 1818 to previously published values, the most
important of these being the preliminary analysis of our entire cluster
sample by Santiago et al.  (2001).  In this section, we compare our
results for the remaining four sample clusters to those of Santiago et
al.  (2001), to illustrate the sensitivity of a simple single-parameter
LF fit to cluster star count data.  The results of this comparison are
presented in Figs.  \ref{slopecf.fig}a, c, d and e.  All of the adopted
luminosity range, radial range, completeness range, and background
subtraction affect an apparently robust result.  For NGC 1831 we also
compare our LF slopes to those obtained from Mateo (1988; Fig. 
\ref{slopecf.fig}b); for NGC 1868 and NGC 2209, Elson et al.  (1999)
published preliminary LFs at various cluster radii based on deep STIS
observations.  We will present a more detailed study of these STIS
observations for all clusters in Beaulieu et al.  (in prep.). 

Since Santiago et al.'s (2001) published annular LF slopes were
determined over a different magnitude fitting range than ours, we
redetermined the slopes for both our LFs and those of Santiago et al. 
(2001) over the maximum common magnitude range available for each of our
clusters (from the MSTO to our 50 per cent completeness limit), after
converting Santiago et al.'s {\sl WFPC2} flight system magnitudes to the
standard {\it V}-band system: (i) for NGC 1831, we used the range $0.0
\le M_V \le 5.1$ ($1.92 \ge \log (L_V / L_{V,\odot}) \ge -0.11$ -- Fig. 
\ref{slopecf.fig}a); (ii) for NGC 1868, $0.65 \le M_V \le 5.1$ ($1.67
\ge \log (L_V / L_{V,\odot}) \ge -0.11$ -- Fig.  \ref{slopecf.fig}c);
(iii) for NGC 2209, $1.3 \le M_V \le 5.1$ ($1.41 \ge \log (L_V /
L_{V,\odot}) \ge -0.11$ -- Fig.  \ref{slopecf.fig}d); and (iv) for Hodge
14, $2.0 \le M_V \le 5.1$ ($1.13 \ge \log (L_V / L_{V,\odot}) \ge -0.11$
-- Fig.  \ref{slopecf.fig}e).  The right-hand subpanels of these figures
show the global LF slopes for each of the clusters; open circles are the
results from our data, while the filled bullets are based on Santiago et
al.'s (2001) published data points, for which we have also indicated the
radial ranges used to obtain the annular LFs by horizontal bars at the
bottom of each figure.  We observe reasonable consistency between our
results, within the associated fitting uncertainties, although a small
discrepancy is seen in the outer regions, beyond $R \simeq 50$ arcsec
for NGC 1868, as for NGC 1805 and NGC 1818 in Paper I.  The comparison
of LF slopes for NGC 2209 from our data and those obtained by Santiago
et al.  (2001) is marginally consistent.  This is most likely due to the
very sparse appearance of this cluster, which renders completeness and
background corrections subject to large uncertainties. 

In Fig.  \ref{slopecf.fig}b, we also compare our LF slopes with those
obtained from the cumulative LF of Mateo (1988).  It is instructive to
see that Mateo's (1988) LF is significantly steeper than ours, which is
most likely due to observational uncertainties, such as crowding,
and blending of sources in his ground-based data. 

\section{Discussion}
\label{tidal.sec}

\subsection{Mass dependence of the core radius -- age relationship}

\begin{figure}
\psfig{figure=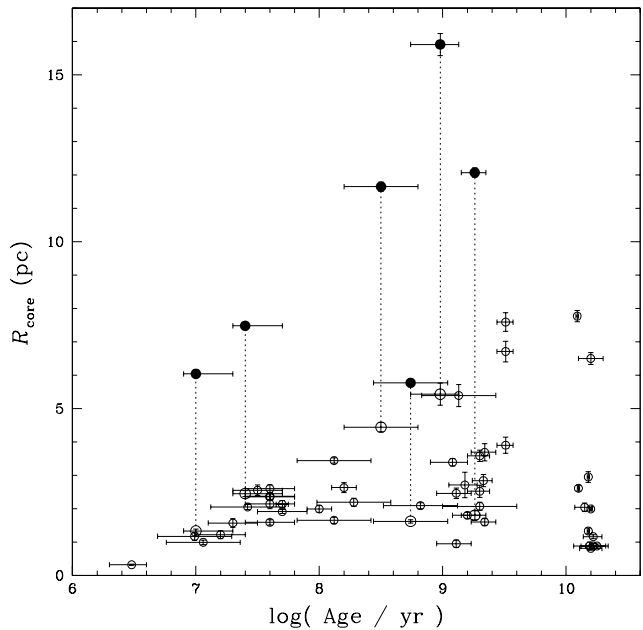,width=9cm}
\caption{\label{lmcclus2.fig}Distribution of cluster core radius versus
age for all LMC clusters in the sample of Mackey \& Gilmore (2002), as
in Fig.  \ref{lmcclusters.fig}.  The vertical dotted lines indicate the
change in core radii if we only consider the cluster stars of $\sim
0.8-1.0 M_\odot$ (filled circles).}
\end{figure}

The reduced spread in core radii for the younger LMC clusters could
possibly be caused artificially if their luminosity profiles are
dominated by a few high-luminosity, high-mass young stars.  This would
be a possible effect of mass-segregated stellar populations, which we
know to be important already for the youngest clusters in our sample
(see Papers I and II).  Such luminosity profiles would not be
representative of the dominant stellar population in the cluster, but
constitute an anomaly.  Therefore, in Fig.  \ref{lmcclus2.fig} we show
the effect of considering only the lowest-mass stars in the
observational LFs of our sample clusters, of mass $\sim 0.8-1.0
M_\odot$.  We see that the increase in upper limit and spread in core
radius with increasing age is retained.  The dynamical relaxation
time-scale for a cluster core radius to grow by the amount observed in
Fig.  \ref{lmcclus2.fig} for the $\sim 1 M_\odot$ stellar population in
all of our clusters is significantly longer than their lifetimes, even
for the oldest clusters (see, e.g., Paper II).  Thus, we conclude that
this effect of increasing upper limit and spread in core radius with age
is real, and that we observe the signature of the initial conditions, at
the time of formation of our sample clusters. 

\subsection{Core expansion due to mass loss and the universality of the
IMF}

A possible explanation offered for the increasing spread in physical
cluster core radii with age between $10^6$ and $10^{10}$ yr (Fig. 
\ref{lmcclusters.fig}) was very large changes in the IMF; a change by 2
in the slope is required.  Cluster-to-cluster variations in the IMF
simulated using Fokker-Planck models (Elson et al.  1989a,b, Elson 1991)
imply different expansion rates of the clusters due to varying mass loss
rates of the evolving stellar population (Chernoff \& Weinberg 1990). 
This rate of expansion of a star cluster is thus governed by its mass
spectrum, i.e., by the MF slope, in the sense that clusters with flat MF
slopes will have formed large fractions of high-mass stars, so that
stellar winds, supernovae and other stellar ejecta would have caused an
important change in the cluster binding energy.  Mass-loss-induced
cluster expansion will continue until a cluster overflows its Roche lobe
and spills beyond its tidal limits, leaving a substantial halo of
unbound cluster stars, which will eventually (on the time-scale of
several orbital periods of the cluster about the LMC, on the order of
$10^9$ yr; see Elson et al.  1987a) be stripped away by the tidal field
of the cluster's parent galaxy (e.g., Elson et al.  1987a, 1989a, van
den Bergh 1991, Goodwin 1997).  This may have partially happened already
in NGC 1831 (Goodwin 1997). 

However, Elson (1991) and Mackey \& Gilmore (2002) point out that,
while local, roughly Salpeter-type IMF variations seem to be able to
explain the small scatter in the core radii around $R_{\rm core} \sim
2.5$ pc, the IMF slopes required to explain the largest core radii --
which are roughly four times larger than the smallest core radii -- are
too flat (IMF slope $\alpha \simeq 0.5$) to allow these clusters to
survive self-disruption beyond $\sim (3-4) \times 10^7$ yr.  In
addition, an increasing body of evidence points towards the universality
of the IMF (see Gilmore 2001 for a review).  Finally, detailed {\it
N}-body simulations by Goodwin (1997), which include the effects of the
expulsion of residual gas from a cluster, and considerations regarding
the initial conditions of cluster formation and the resulting star
formation efficiency (see also Elson et al.  1987a), also appear to be
unable to produce a fourfold increase in the cluster core radii over
their lifetimes, even if the star formation efficiency remains low
throughout. 

As shown in Fig.  \ref{comparelf.fig}, both the change of the LF (or MF)
slope with radius, and the absolute LF (MF) slopes between the two
clusters in each of our cluster pairs of similar age are identical
within the observational (and systematic) uncertainties.  If there is
such a thing as a universal IMF, then a minimal expectation would be
that the younger LMC clusters, with no significant dynamical evolution,
a wide range of stellar masses, and in some cases very similar
metallicities, should have indistinguishable mass functions.  This is
supported by the very similar LF (MF) slopes as a function of radius for
NGC 1805 and NGC 1818. 

Moreover, Fig.  \ref{compareall.fig} shows unambiguously, that this
result also holds for {\it all} clusters in our sample, irrespective of
their evolutionary state or core radius: {\it both} the central LF slope
{\it and} the degree of mass segregation, as seen from the gradient of
the LF slope with radius, are confined within narrow ranges, at most
spanning a $\sim (2-3) \sigma$ range in parameter space.  While the
intermediate result in Fig.  \ref{comparelf.fig} merely allowed us to
conclude that the present-day MFs of the clusters in each pair must be
very similar, the similarity of the degree of mass segregation and of
the actual LF slopes in the inner cluster regions among all of our
sample clusters implies that their {\it initial} MF must have been very
similar, if not identical, within the uncertainties: after all, if the
IMF in each cluster had been different, it seems highly unlikely that
the MFs as a function of (core) radius of all of our clusters are
currently so similar.  This is yet another important result in favour of
a universal IMF among star clusters of widely disparate properties. 

Similarly, Elson et al.  (1999) concluded from a pilot study of the LFs
of NGC 1868 and NGC 2209 based on deep {\sl HST}/STIS observations, that
IMF variations do not appear to be responsible for the differences in
core radii between these clusters. 

Although we are strongly in favour of using LFs instead of their
associated MFs, due to the large systematic uncertainties involved in
the luminosity-to-mass conversions, we can still reach robust
conclusions on the importance of the steepness of the MF slopes with
respect to the Salpeter IMF slope for our sample clusters.  In Paper II
we converted the LFs of the two youngest clusters in our sample, NGC
1805 and NGC 1818, to present-day MFs.  Although the effects of mass
segregation in these clusters resulted in a radial dependence of the MF
slope, the overall cluster slopes were found to be close to the generic
Salpeter IMF slope, or perhaps slightly steeper, depending on the mass
fitting range and luminosity-to-mass conversion used (for $-0.15 \le
\log m/M_\odot \le 0.85$; see Paper II).  The result visualized in Fig. 
\ref{compareall.fig} indicates that the global present-day MF slopes for
NGC 1831, NGC 1868, NGC 2209 and Hodge 14 are also similar to each other
at masses down to $\sim 0.8 - 1.0 M_\odot$, and are certainly not
flatter than the higher-mass IMF slope.  If we then assume that the main
effect of mass segregation is a redistribution of the cluster stars,
this implies that the {\it initial} MF of all clusters is closely
represented by a Salpeter IMF. 

This is consistent with the result of Mateo (1988), who concluded that
the IMF slopes in the range from $0.9 - 10.5 M_\odot$ for six young and
intermediate LMC star clusters, including NGC 1831, are remarkably
similar to the Scalo (1986) IMF for field stars in the solar
neighbourhood (see also Sagar \& Richtler [1991], and Banks, Dodd \&
Sullivan [1995] for Salpeter-like MF slopes in LMC clusters).  

The full range of variation in IMF slope allowed by our observations is
a few tenths.  By contrast, the models require an order of magnitude
larger change in slope to explain the core evolution.  We conclude that
IMF variations in our sample clusters do not drive the core radius--age
relationship. 

\subsection{The effects of a significant binary population}

Alternatively, it has been argued that the spread in core radii towards
greater ages might be due to the effects of a very large difference in
the (unresolved) binary and multiple star population in these clusters. 
For young star clusters the optical/near-infrared CMDs often show a
clear binary sequence parallel to the single-star main sequence (Hut et
al.  1992, Elson et al.  1998, Johnson et al.  2001), for older clusters
the main sequence becomes almost vertical, thus hiding a possible binary
sequence. 

For the LMC clusters in our sample, Chiosi (1989) argues that assuming a
30 per cent binary fraction for NGC 1831 yields a perfect match to the
observed broadening of the CMD, in particular of the red giant branch. 
Elson et al.  (1998) show for NGC 1818, that the binary fraction
increases towards the cluster centre from $\sim (20 \pm 5)$ per cent in
the outer parts to $\sim (35 \pm 5)$ per cent inside the core, with mass
ratios $\gtrsim 0.7$.  They argue that this increase is entirely
consistent with predicted dynamical mass segregation effects, based on
{\it N}-body calculations. 

Binary stars play a dynamically important role in the evolution of star
clusters (Elson, Hut \& Inagaki 1987b, Hut et al.  1992, Meylan \&
Heggie 1997).  Observations seem to show that all clusters have a
substantial binary fraction, and that large cluster-to-cluster
variations are not found.  It is unlikely that the allowed variations of
up to a factor of two can produce the roughly fourfold increase in core
radii observed between the youngest and the oldest LMC clusters. 

\subsection{Merging binary clusters?}

Finally, {\it N}-body simulations of encounters betweeen unequal-mass
clusters (e.g., Rao, Ramamani \& Alladin 1987, Barnes \& Hut 1986, 1989,
Rodrigues et al.  1994, de Oliveira, Dottori \& Bica 1998) have shown
that external effects, such as mergers and tidal disruption, are
important processes in the dynamical evolution of binary clusters. 

The observational evidence for (i) bumps, sharp shoulders and central
dips in the radial surface brightness profiles of young and
intermediate-age LMC clusters (including NGC 1818, e.g., Elson et al. 
1987a, Elson 1991, Mackey \& Gilmore 2002), (ii) the markedly non-zero
ellipticity of some clusters with large core radii (e.g., NGC 1818 and
NGC 1831; Sugimoto \& Makino 1989, Elson 1991, de Oliveira et al. 
2000), and (iii) the evidence for a constant or declining star formation
rate, or -- alternatively -- multiple bursts of star formation within a
single star cluster (e.g., Chiosi 1989) have led to the suggestion that
these could simply be manifestations of merging binary (sub)clusters
(e.g., Elson 1991, de Oliveira et al.  2000) or subunits within a single
progenitor molecular cloud complex (e.g., Bhatia \& MacGillivray 1988,
Elson 1991).  It is well known that interacting binary, presumably
coeval cluster pairs are fairly common in the LMC (e.g., Bhatia, Cannon
\& Hatzidimitriou 1987, Bhatia \& Hatzidimitriou 1988, Bhatia \&
MacGillivray 1988, Bhatia et al.  1991, Bica \& Schmitt 1995, de
Oliveira et al.  1998, 2000, Bica et al.  1999; see also Mackey \&
Gilmore 2002), which could merge on relatively short time-scales given
suitable conditions (Bhatia 1990), possibly leading to significant core
expansion.  De Oliveira et al.  (2000) show, based on {\it N}-body
modeling, that the merger of an unequal-mass binary cluster pair can
reach a stable state on time-scales as short as $\sim 200$ Myr, after
which it can have attained a significantly different structure and
ellipticity from the original main cluster. 

NGC 1831 is one of the clusters with the largest core radii in our
sample.  De Oliveira et al.  (2000) show that its structure is
consistent with models of a merged system of binary clusters.  In
addition, NGC 1831 and NGC 1868 have peculiar CMDs (e.g., Chiosi 1989,
Santiago et al.  2002), which can be interpreted as (i) the result of
constant, continuously declining, or multiple bursts of star formation
(e.g., Chiosi 1989), or (ii) having been caused by a merger of a binary
cluster system (e.g., Chiosi 1989, Santiago et al.  2002): although
binary clusters are likely fairly coeval, age differences of $\sim 10^7
- 10^8$ yr are not ruled out (Elson et al.  1987a, Chiosi 1989, and
references therein). 

De Oliveira et al.'s (2000) {\it N}-body simulations show that the
merger of two clusters with a mass ratio of 10:1, with the less massive
one orbiting the massive cluster on an elliptical orbit, will eventually
lead to the disruption of the smaller cluster.  The end product of such
a merger is a single cluster, with the stars of the disrupted cluster
forming a halo around the final cluster in the original orbital plane of
the less massive cluster, while some are ejected from the system. 
Typically, in the absence of an external tidal field, $\lesssim 50$ per
cent of the mass of the disrupted cluster member ($\lesssim 4.5$ per
cent of the total mass of the system) will be dispersed into the field,
i.e.  beyond the tidal truncation radius of the final cluster. 

However, the number of true binary cluster candidates is far too small
for this scenario to be very important.

\section{Summary and Conclusions}

In this paper, we have quantified mass segregation as a function of
cluster core radius in a sample of LMC clusters in order to investigate
the trend of the upper limit on the core radius of the LMC cluster
system to increase with increasing cluster age.  We discuss the
implications of the observed degree of mass segregation for the shape of
the IMF. 

Our results are based on {\sl HST} observations of six rich star
clusters in the LMC, selected to include three pairs of clusters of
similar age, mass, metallicity, and projected distance from the LMC
centre, while we required the largest possible spread in core radii
between the clusters in each pair. 

We study the dependence of the shape and slope of the stellar LF on
position within the clusters.  Although there is a one-to-one
correlation between a cluster's LF and its associated MF, in view of the
uncertainties involved in the conversion of luminosities to masses we
{\it only} used the LFs to reach our conclusions on the effects of mass
segregation.  This approach is therefore less model-dependent and leads
to identical, robust results, without having to keep in mind the large
systematic uncertainties inherent to any luminosity-to-mass conversion. 

All of our sample clusters show clear evidence of mass segregation, in
the sense that (i) the LF (MF) slopes steepen with increasing cluster
radius, and (ii) the brighter stars are increasingly concentrated
towards the cluster centres (i.e., they are characterized by smaller
core radii): while the effects of mass segregation are most clearly seen
for stellar masses $\log m/M_\odot \gtrsim 0.2$ in the youngest sample
clusters, NGC 1805 and NGC 1818, clear mass segregation is seen for the
four older star clusters down from the MSTO magnitudes. 

Although the trend towards steeper LFs with increasing radius is clear,
the associated error bars are large, clearly reflecting the
non-linearity of the annular LFs.  Nevertheless, it is apparent from the
comparison of the dependence of the LF slope as a function of radius
between the clusters in each pair, that, within the uncertainties, this
dependence is identical for both clusters in a given age pair.  If there
is such a thing as a universal IMF, then a minimal expectation would be
that the younger LMC clusters, with no significant dynamical evolution,
a wide range of stellar masses, and in some cases very similar
metallicities, should have indistinguishable mass functions.  This is
supported by the very similar LF (MF) slopes as a function of radius for
NGC 1805 and NGC 1818. 

Moreover, we find that {\it both} the central LF slope {\it and} the
degree of mass segregation of our sample clusters, as indicated by the
gradient of the LF slope with radius, are confined within narrow ranges,
at most spanning a few tenths in slope.  This result is indeed rather
surprising in view of the large range in age (and therefore in dynamical
state), mass, metallicity and structural parameters (core radii)
occupied by the ensemble of our sample clusters.  While the intermediate
comparison of the LFs of clusters within a given pair allowed us to
conclude that the present-day MFs of the clusters in each pair must be
very similar, the similarity of the degree of mass segregation and of
the actual LF slopes in the inner cluster regions among all of our
sample clusters implies that their {\it initial} MF must have been very
similar, if not identical, within the uncertainties, down to masses of
$\sim 0.8 - 1.0 M_\odot$ (depending on the luminosity-to-mass conversion
adopted).  This is yet another important result in favour of a universal
IMF among star clusters of widely disparate properties. 

We can thus firmly rule out variations in the IMF as the main driver of
the increasing spread of cluster core radii as a function of increasing
age in our cluster sample.  We are currently investigating the
evolutionary effects of the LMC tidal field on its star cluster
population.  The results of this analysis, based on $N$-body
simulations, will be published elsewhere (Wilkinson et al., in prep.). 

\section*{Acknowledgments} This paper is based on observations with the
NASA/ESA {\sl Hubble Space Telescope}, obtained at the Space Telescope
Science Institute, which is operated by the Association of Universities
for Research in Astronomy (AURA), Inc., under NASA contract NAS 5-26555. 
This research has made use of NASA's Astrophysics Data System Abstract
Service.

\end{document}